\documentclass[prc,aps,twocolumn]{revtex4}
\usepackage{graphicx}
\usepackage{color}
\begin{document}

\title{ The radius of a typical-mass neutron star and chiral effective field theory      
} 

\author{F. Sammarruca and Randy Millerson}
\affiliation{Department of Physics, University of Idaho, Moscow, ID 83844, USA}


\begin{abstract}
We calculate neutron star masses and radii from equations of state based on recent 
high-quality chiral nucleon-nucleon
potentials up to fifth order of 
the chiral expansion and the leading 
chiral three-nucleon force. Our focus is on the radius of a 1.4 $M_{\odot}$ neutron star, for 
which we report predictions that are 
consistent 
with the most recent constraints. We also show the
full $M(R)$ relations up to their 
respective maximum masses.                                                                 
Beyond the densities for which microscopic predictions are derived from chiral forces,                 
the equations of state are obtained {\it via} polytropic continuations. 
However, the radius 
of a 1.4 $M_{\odot}$ neutron star is nearly insensitive to the high-density extrapolation.               

\end{abstract}
\maketitle 

\section{Introduction} 
\label{Intro} 

Neutron-rich systems are associated with a variety of important and still open        
questions such as: the location of neutron drip lines, the thickness of neutron skins, and the structure 
of neutron stars. 
Common to these diverse situations is the equation of state (EoS) of neutron-rich matter, 
namely the energy per particle in isospin-asymmetric matter as a function of density 
(and other thermodynamic quantities if appropriate, such as temperature). In the presence of different
neutron and proton concentrations, the symmetry energy emerges as an important component of the EoS  
and plays an outstanding role in the physics of neutron-rich systems. 
In fact, it is remarkable that 
the relation between the mass and the radius of neutron stars is uniquely determined
by the EoS together with their self-gravity. In short, these compact systems are intriguing testing grounds for 
nuclear physics. 
Most recently, the detection by LIGO of 
gravitational waves from two neutron stars spiraling inward and merging has generated even 
more interest and excitement 
around these exotic systems. 
In fact, the LIGO/Virgo~\cite{merg1} detection of gravitational waves originating from the neutron star merger GW170817
has most recently provided new and more stringent constraints on the maximum radius of a 1.4 $M_{\odot}$ 
neutron star, based on the tidal deformabilities of the colliding stars~\cite{merg2}. 

A main purpose of this paper is to present and discuss predictions of neutron star masses and radii  
based, as far as possible, on state-of-the-art nuclear forces. 
The focal point is the radius of a star with mass equal to 1.4 $M_{\odot}$ (the
most probable mass of a neutron star), which we wish to predict
with appropriate quantification of the theoretical error.

When obtained microscopically, the EoS results from the application of few-nucleon forces in appropriate 
many-body calculations, such as, for instance, the Brueckner-Hartree-Fock approach to 
nuclear matter. This is in contrast to 
methods which obtain the EoS from phenomenological functionals parametrized in such a 
way as to describe selected nuclear 
properties. 

Concerning the development of modern few-nucleon forces, 
recently chiral Effective Field Theory (EFT) has become established as               
the most fundamental and potentially 
model independent approach to nuclear forces. Chiral EFT                                          
is firmly based on the symmetries of low-energy quantum chromodynamics (QCD) and, furthermore, 
its predictions can be improved in a systematic way~\cite{ME11,EHM09}. This is not the case with forces developed,
for instance, within the formerly very popular meson-exchange picture~\cite{Mac89,AV18}.
Furthermore, the problem with the meson-exchange model is that three-nucleon forces (3NF) or, more generally, $A$-nucleon forces with $A>2$ do not have a firm link with the two-nucleon force (2NF) to which they are associated.

Chiral EFT, however, is a low-energy theory and thus there are
limitations to its domain of applicability. First, the chiral symmetry breaking scale,
$\Lambda _{\chi} \approx 1$ GeV, sets a limit to the momentum or energy domains where pions and nucleons are
the appropriate degrees of freedom. Moreover, the cutoff parameter $\Lambda$ appearing in the regulator function 
suppresses high momentum components, which should play no role
in the prediction of low-energy observables, to a degree which depends on the strength of the cutoff.

Central densities in compact stars can exceed several times the density of normal saturated matter,       
and of course high densities imply the presence of             
high Fermi momenta.                  
Those are outside the reach of chiral EFT and therefore 
methods to extend the predictions must
be employed. A reasonable guidance for how to extrapolate chiral predictions to high densities may be the 
consideration that, for a very large number of existing EoS, the pressure    
as a function of baryon density 
(or mass density) can be fitted by piecewise polytropes, namely functions of the form 
$P= \alpha \rho ^{\Gamma}$~\cite{poly}. (In our notation, $\rho$ denotes the baryon density.)
With this observation as a guideline, 
after outlining       
our calculations of the energy density and pressure for matter with 
nucleons and leptons in $\beta$ equilibrium, 
see Sections~\ref{chi}-\ref{bet},                                                                       
we will extend the pressure predictions obtained from the chiral EoS                                                  
using polytropes, see 
Section~\ref{cont}.                  

 We are then in the position to solve the      
Tolman-Oppenheimer-Volkoff (TOV)~\cite{tov}        
star structure equations to obtain the mass as a function of the radius for a sequence of stars differing in their
central densities, up to several times normal density. We will consider             
stellar matter with neutrons, protons, and leptons in 
$\beta$ equilibrium.                                         

 Extensive effort has been dedicated to constraining properties of compact stars from astrophysical observations,
see, for instance, Ref.~\cite{LP04,LP07,LP16,SLB13,Oz13}.
We will compare our predictions with the most recent constraints from astrophysical data.          
A brief summary, conclusions, and future plans are contained in 
Sect.~\ref{Conc}.

\section{The Chiral forces}                                  
\label{chi} 

As mentioned in the Introduction, 
at this time chiral EFT is the only path for constructing              
 nuclear two- and many-body forces in a systematic and, ideally, model-independent 
way~\cite{ME11,EHM09}.                                     

Chiral nucleon-nucleon (NN) potentials have been made available from leading order (LO, zeroth order) to
N$^3$LO (fourth order)~\cite{ME11,EHM09,NLO,EM03,EGM05}, with the latter reproducing NN data with high precision.
 More recently, chiral NN potentials at N$^4$LO have also become available~\cite{n4lo,EKM15}.

Chiral interactions have been applied in few-nucleon 
reactions~\cite{Epe02,NRQ10,Viv13,Gol14,Kal12,Nav16}, structure of light- and medium-mass nuclei~\cite{Coraggio07,Coraggio10,Coraggio12,Hag12a,Hag12b,BNV13,Gez13,Her13,Hag14a,Som14,Heb15,Hag16,Car15,Her16,Hol17,Sim17,Mor17},
infinite matter at zero temperature~\cite{HS10,Heb11,Baa13,Hag14b,Cor13,Cor14,Sam15,Dri16,Tew16,Hol17}
and finite temperature~\cite{Wel14,Wel15}, as well as other aspects of nuclear dynamics~\cite{Bac09,Bar14,Rap15,Bur16,Hol16,Bir17,Rot17}.

\subsection{The two-nucleon forces}                                  
\label{2nchi} 

Next, we briefly summarize the main features of the 2NFs employed in this work. 
The reader is referred to Ref.~\cite{n4lo} for a complete and detailed description. 
         
The $NN$ potentials span five orders of chiral EFT, from leading order (LO) to fifth order (N$^4$LO). 
The same power counting scheme and regularization procedures are applied through all orders, making this set 
of interactions more consistent than previous ones. 

Another novel and important aspect in the construction of these new potentials is the fact that the long-range 
part of the interaction is fixed by the $\pi N$ LECs as determined in the recent and very accurate analysis of Ref.~\cite{Hofe}. In fact, for all practical purposes, errors in the $\pi N$ LECs are no longer an issue with regard to uncertainty quantification.
Furthemore, at the fifth (and highest) order, the $NN$ data below pion production threshold are reproduced with 
excellent precision ($\chi ^2$/datum = 1.15).

Iteration of the potential in the Lippmann-Schwinger equation requires                         
cutting off high momentum components, consistent with the fact that chiral perturbation theory amounts to building a low-momentum expansion.          
This is accomplished through the application of a regulator function for which the non-local form                                
\begin{equation}
f(p',p) = exp[-(p'/\Lambda)^{2n} - (p/\Lambda)^{2n}]     
\label{reg}
\end{equation}
is chosen.
For the present applications in nuclear and neutron matter, we will limit ourselves to values of the 
cutoff parameter $\Lambda$ smaller than or equal to 500 MeV, as those have been associated with the onset of favorable perturbative properties~\cite{FS+18}.

\subsection{ The three-nucleon forces} 
\label{3nchi} 

Three-nucleon forces make their first appearance at the third order of the chiral expansion (N$^2$LO). At this order, the 3NF consists of three contributions~\cite{Epe02}:
the long-range two-pion-exchange (2PE) graph, 
the medium-range one-pion exchange (1PE) diagram, and a short-range contact term. 
We apply these 3NFs by way of the density-dependent effective two-nucleon interactions derived in 
Refs.~\cite{HKW}. They are expressed in terms of the well-known 
non-relativistic two-body nuclear force operators and can be conveniently 
incorporated in the usual $NN$ partial wave formalism and the conventional Brueckner-Hartree-Fock theory.

The LECs $c_D$ and $c_E$ which appear within the three-nucleon sector are             
constrained so as to reproduce the $A=3$ binding energies and the Gamow-Teller matrix element of tritium $\beta$-decay, following an established procedure which has been recently revisited in Ref.~\cite{Marc18}.

The complete 3NF at orders higher than three is very complex, which is why including only the leading 3NF is a common practice. 
However, there is one important component of the 3NF where complete
calculations up to N$^4$LO are in fact possible, namely the 2PE 3NF. In Ref.~\cite{KGE12} it has been
shown that the 2PE 3NF mathematical structure is nearly the same at N$^2$LO, N$^3$LO, and N$^4$LO.
Thus, the three orders of 3NF contributions can be added up and parametrized in terms of effective 
$c_i$ LECs. We will follow this scheme in the present calculations. The reader is referred to 
Refs.~\cite{FS+18} for a detailed description of the EoS based on the newest potentials~\cite{n4lo}.

We note that, although many 
3NF contributions are possible, the 2PE 3NF is of paramount significance (and the first             
3NF to be calculated~\cite{FM57}). The prescription briefly outlined above allows to include 
this very important 3NF component up to the highest orders being considered.

\section{Calculation of the EoS in $\beta$-equilibrated matter}
\label{bet}

We calculate the EoS of symmetric nuclear matter and pure neutron matter and thus the symmetry energy.
The calculations are performed with the interactions as described in the previous section. We use the 
non-perturbative particle-particle ladder approximation, which generate the leading order in the traditional
hole-line expansion. 

\begin{figure}[!t] 
\centering
\vspace*{-1cm}
\includegraphics[width=8.7cm]{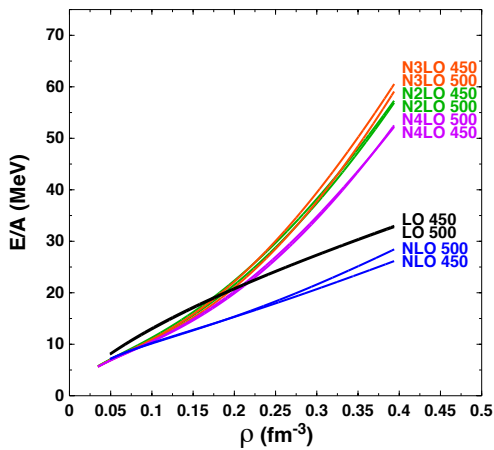}
\vspace*{-3cm}
\caption{Energy per particle in neutron matter as a function of density at the indicated orders and
with varying cutoff parameters as denoted.                                       }    
\label{fig1}
\end{figure}

As a demonstration of the order-by-order pattern in our latest EoS~\cite{FS+18},
we show in Fig.~\ref{fig1} the energy per particle in neutron matter across all five orders of the 
chiral perturbation expansion and for two 
values of the cutoff parameter. Note that no 3NF are present at leading and next-to-leading orders.
Irrespective of that, NN data cannot be described at a satisfactory precision level below the third order.
Therefore,
in what follows we will        
show predictions only at realistic orders, namely equal or above the third (N$^2$LO).

The total energy per particle, $e_{tot}$, 
related to the total energy density, $\epsilon_{tot}$, by                                                
$ e_{tot} = \epsilon_{tot}/\rho$, for neutrons and protons in 
$\beta$ equilibrium with electrons and muons is given by:
\begin{equation}
e_{tot} = e_0 + e_{sym} (Y_n-Y_p)^2 +      
 \sum_{i=n,p} Y_i m_i  +      
 e_e  + e_\mu   \; .                            
\label{etot}
\end{equation} 
The first three terms on the right-hand side are the baryon contributions with their rest energies 
($Y_i$, $i=n(p)$, stands for the neutron(proton) fraction), 
while the last two are the relativistic electron and the 
muon energies per baryon.                                                                       
We then proceed to minimize the total energy per particle subjected to the constraints 
of fixed baryon density and charge neutrality. 
The resulting set of equations allow to solve for the various lepton fractions and then obtain        
the corresponding 
energy densities.

\begin{figure*}[!t] 
\centering
\vspace*{-1cm}
\includegraphics[width=6.7cm]{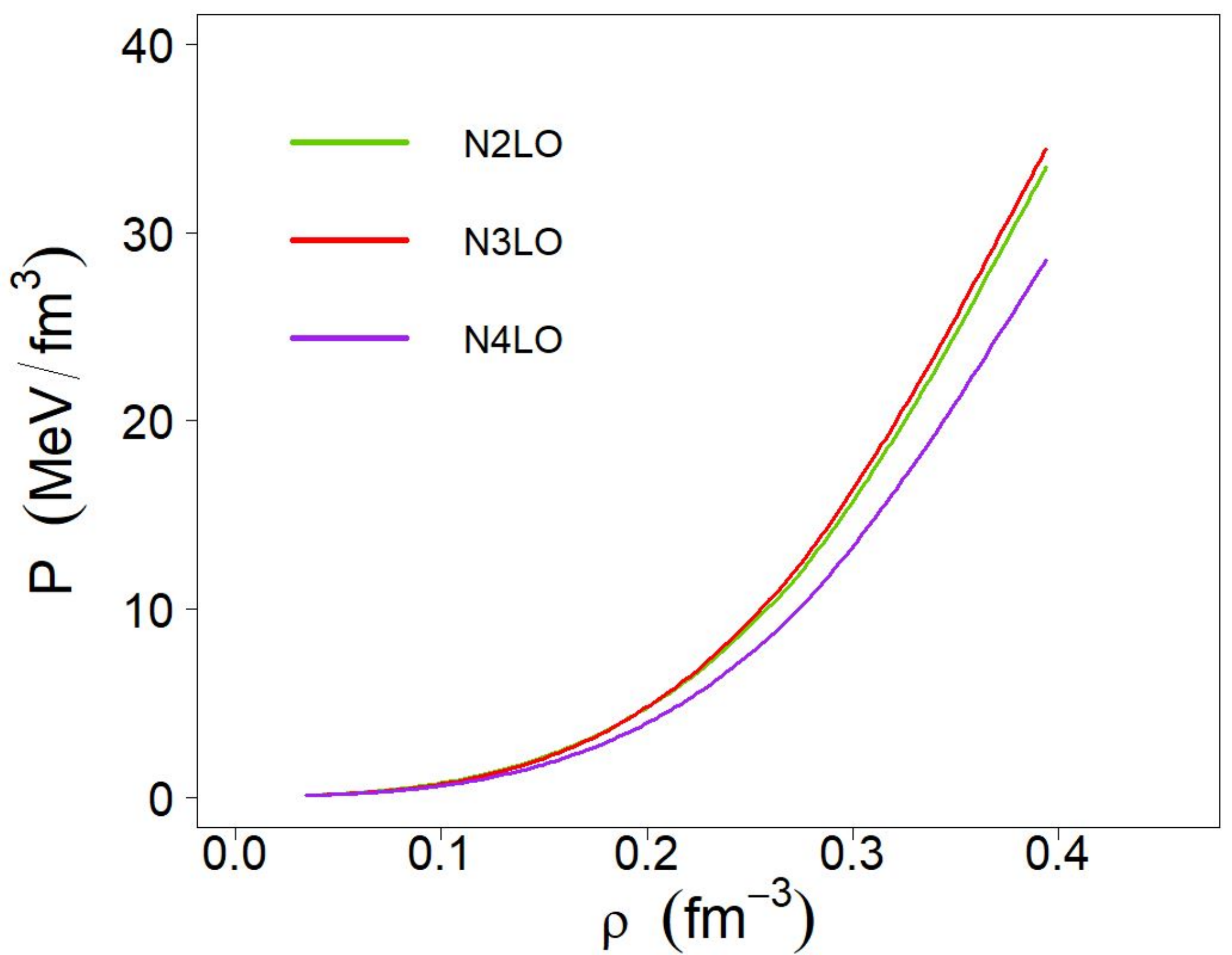}\hspace{0.01in} 
\includegraphics[width=6.7cm]{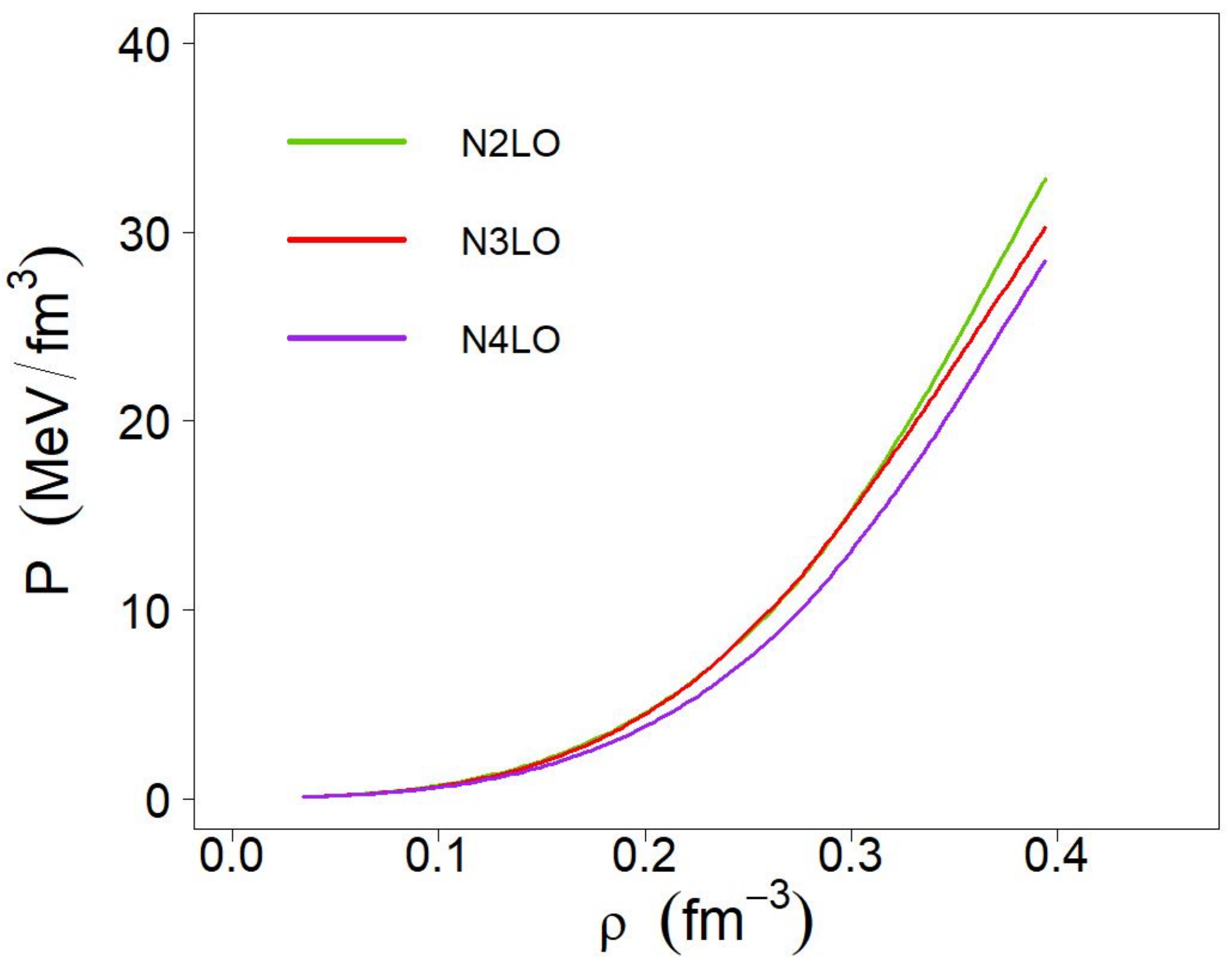}
\vspace*{1.0cm}
\caption{Pressure in $\beta$-stable matter as a function of density at the indicated orders for $\Lambda$=450 MeV (left) and $\Lambda$=500 MeV (right). 
}                                                            
\label{pbands}
\end{figure*}

The pressure is related to the energy density through
\begin{equation}
P(\rho) = \rho^2 \frac{d (\epsilon_{tot}/\rho)}{d \rho} \; .                       
\label{pres} 
\end{equation}
In Fig.~\ref{pbands}, we show the calculated pressure in $\beta$-stable matter at the third, fourth, and fifth 
orders of the 2NF together with the leading 3NF.           

\section{Continuing the EoS to high density}                                  
\label{cont}

In this section, we perform continuation to high-density of the microscopic EoS. 
We employ our microscopic predictions up to about 2$\rho_0$, a choice which calls
for some explanations.                              
Since we are dealing with a perturbative expansion in the parameter $Q/\Lambda$, we rely on                   
arguments based on the size of the expansion parameter for typical momenta of the system under
consideration.                                             
Smooth regulators can, of course, impact momenta lower than the highest 
momentum, which, for 
neutron-rich matter                                              
around twice normal           
density (with our predicted proton fractions), is approximately 
400 MeV, still lower 
than, although close to $\Lambda
\sim 450-500$ MeV. On the other hand, the {\it r.m.s.} momentum of nucleons in nuclear matter is approximately 
55\% of the Fermi momentum~\cite{HT}.                                                  
Thus, on statistical grounds, we should be safe from cutoff artifacts.

We then attach polytropes having different 
adiabatic indices, $P(\rho) = \alpha \rho ^{\Gamma}$, 
ensuring, of course, continuity 
of the pressure.    
The range of the polytropic index is taken to be between 1.5 and 4.5 (based on guidelines from 
the literature~\cite{poly}), and these extensions are calculated up to about 3$\rho_0$. At this density, 
every polytrope is again joined continuosly with a set of polytropes spanning the same range.                   
In this way, we can cover a large set of possibilities, with the EoS being 
``softer" or ``stiffer" in one density region or the other, as it would be the case if phase transitions 
(most likely to non-hadronic degrees of freedom) were to take place. 

\begin{figure*}[!t]
\includegraphics[width=7.0cm]{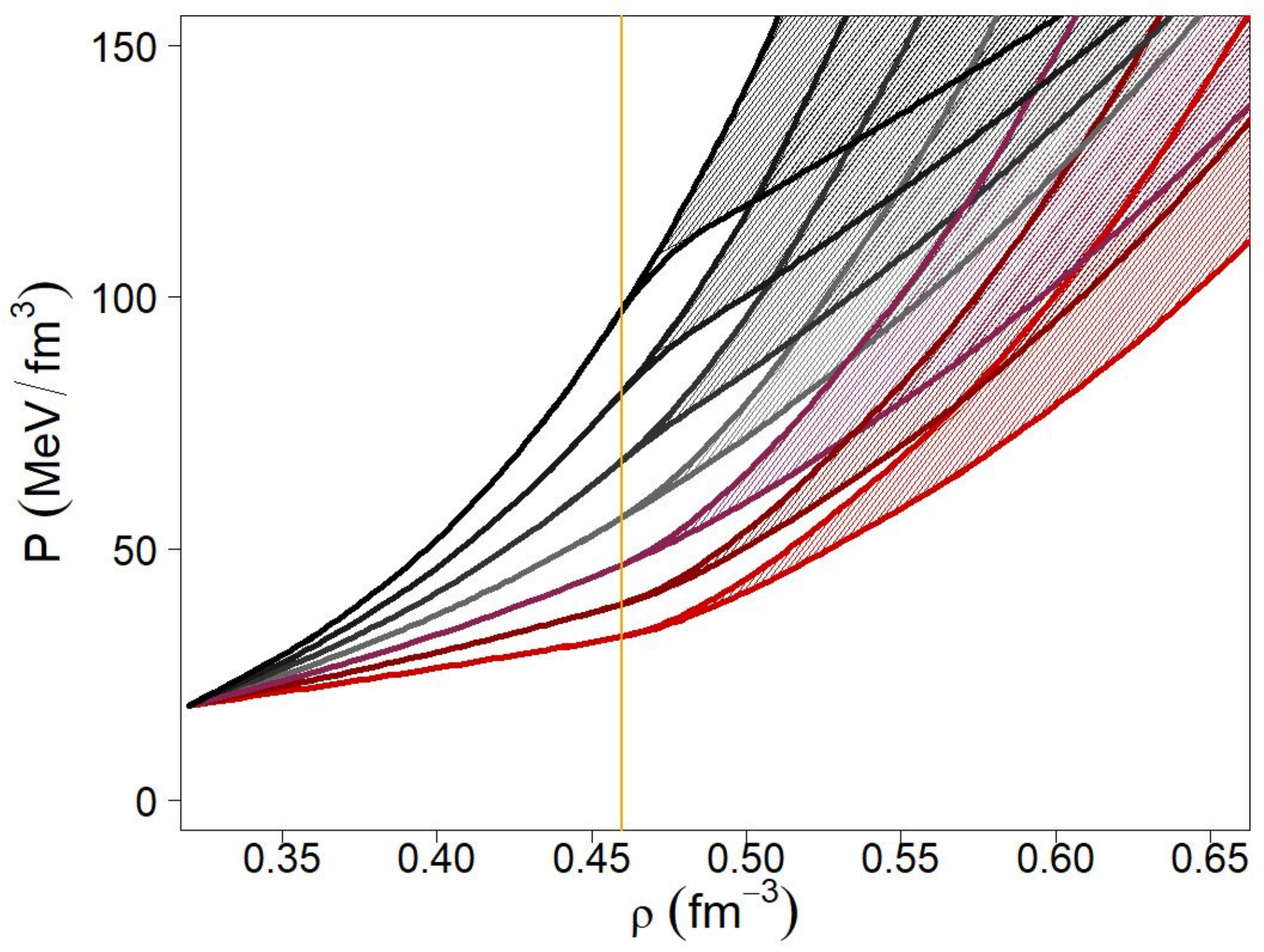}
\includegraphics[width=7.0cm]{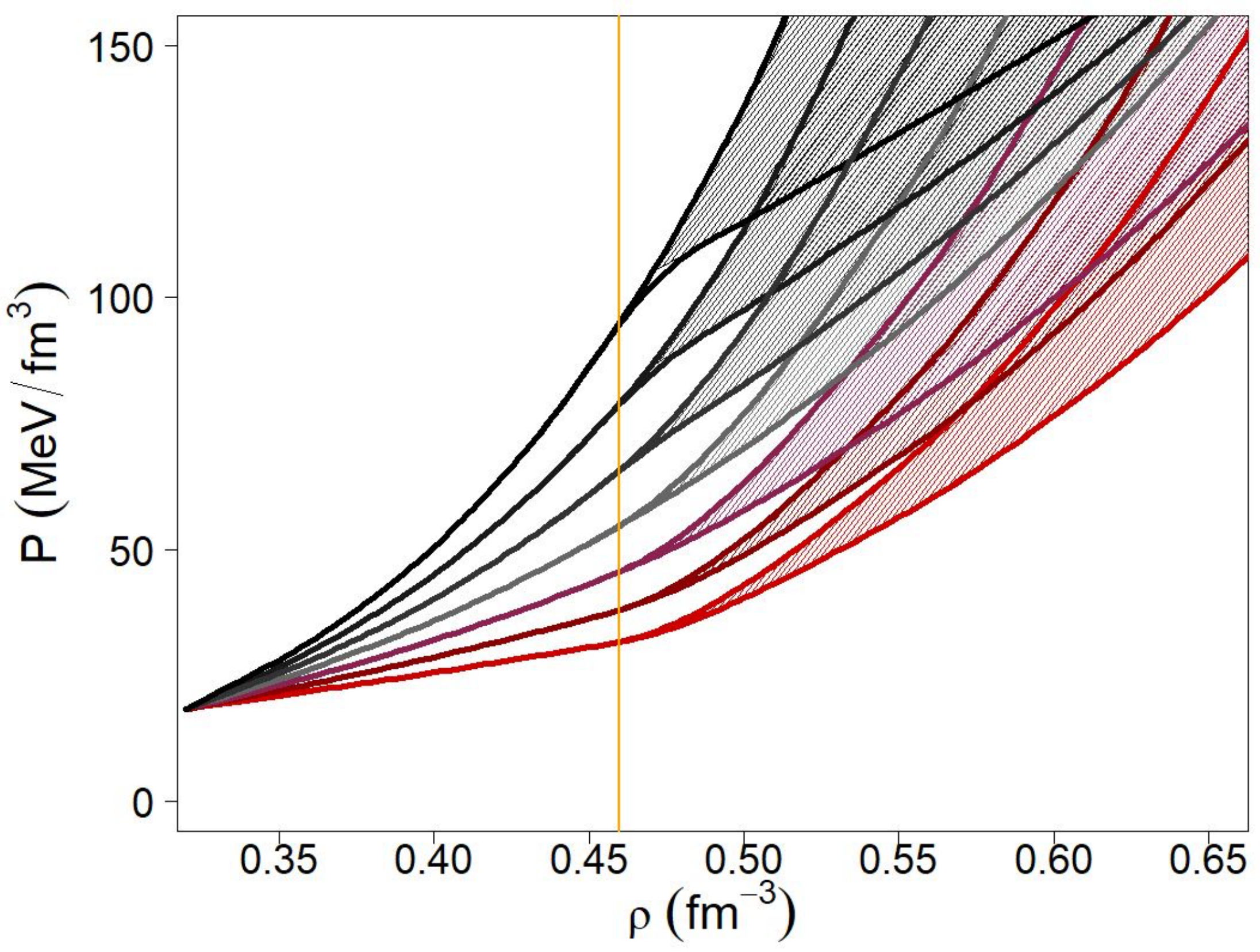}
\vspace*{0.5cm}
\includegraphics[width=7.0cm]{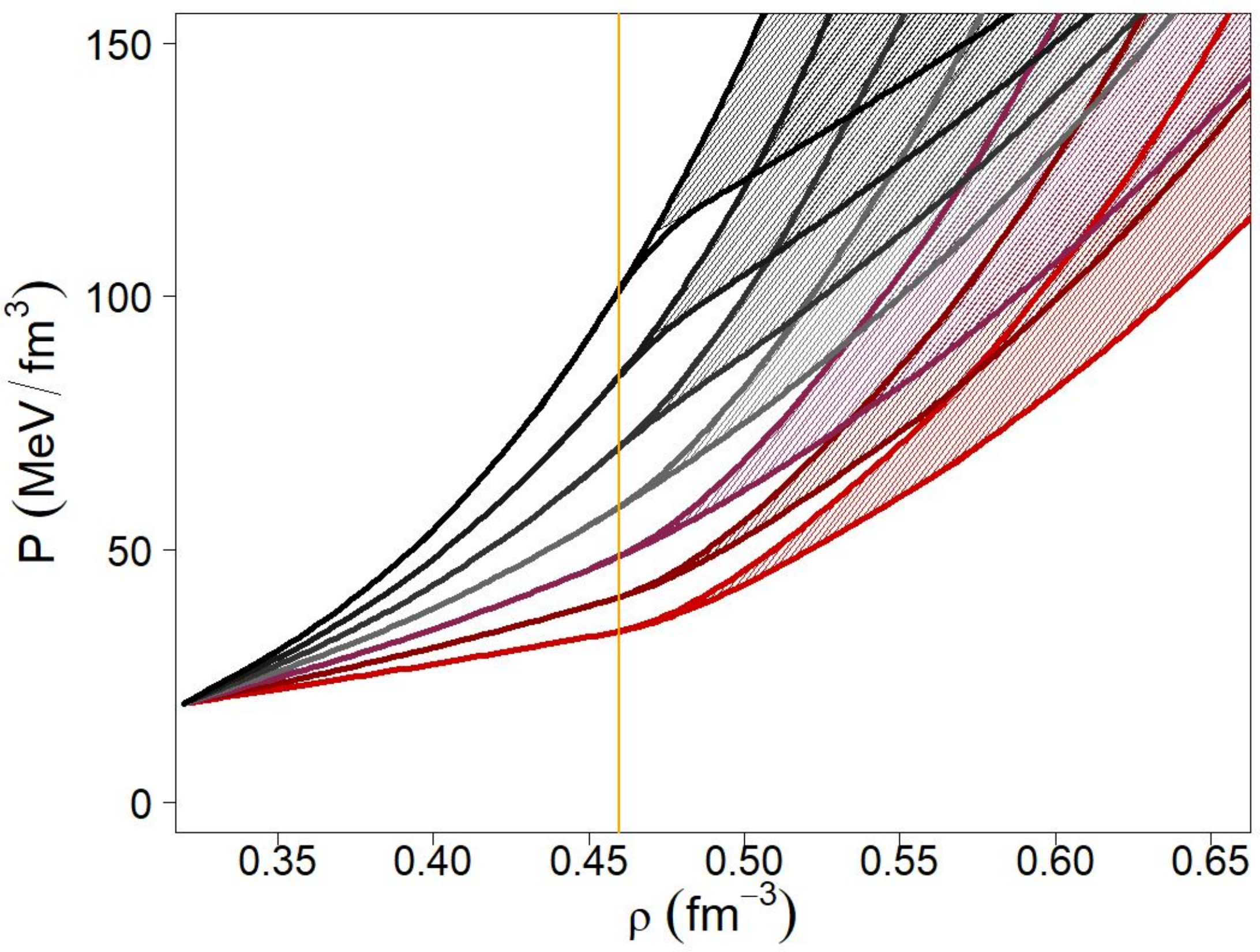}
\includegraphics[width=7.0cm]{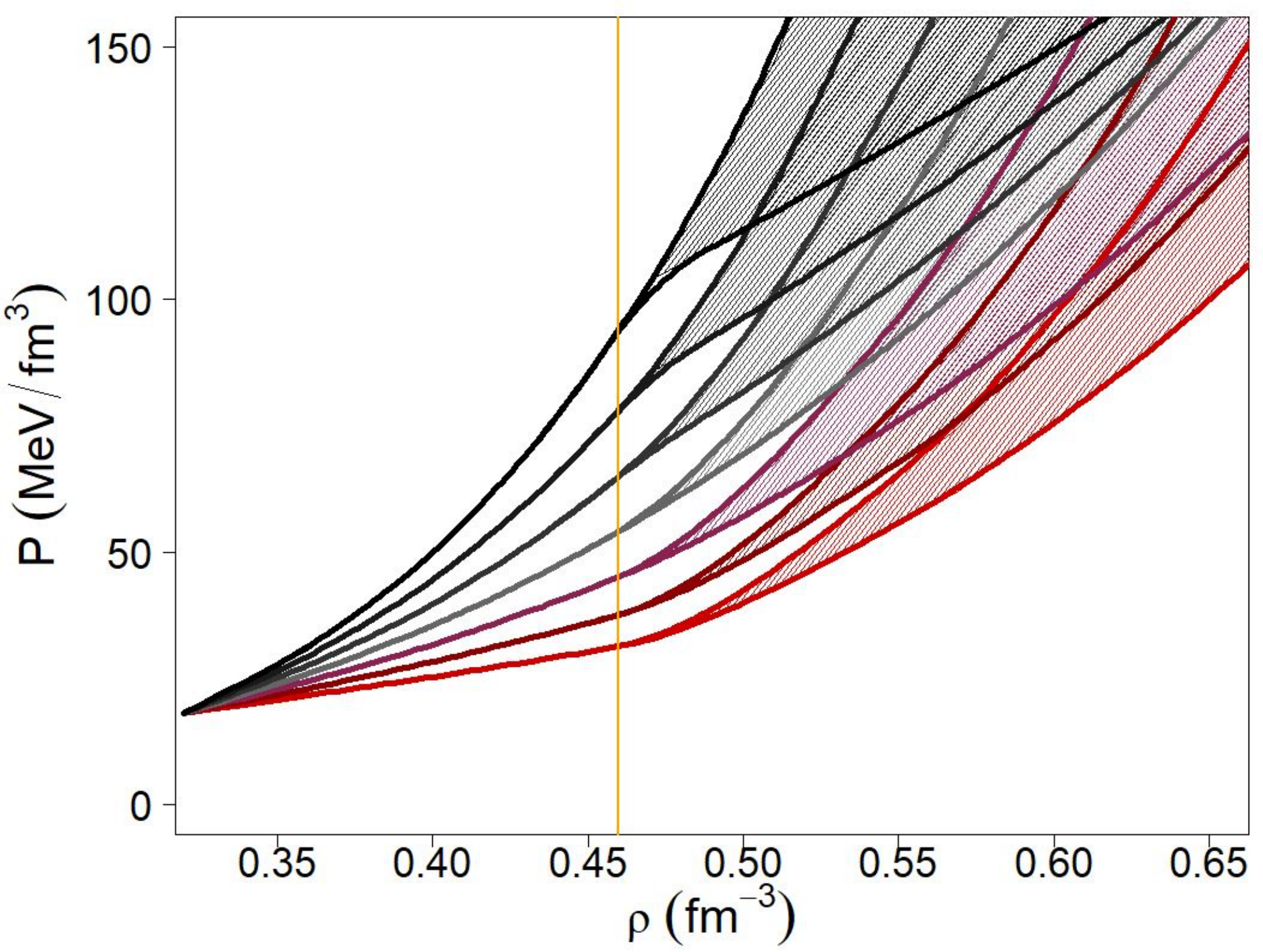}
\vspace*{0.5cm}
\includegraphics[width=7.0cm]{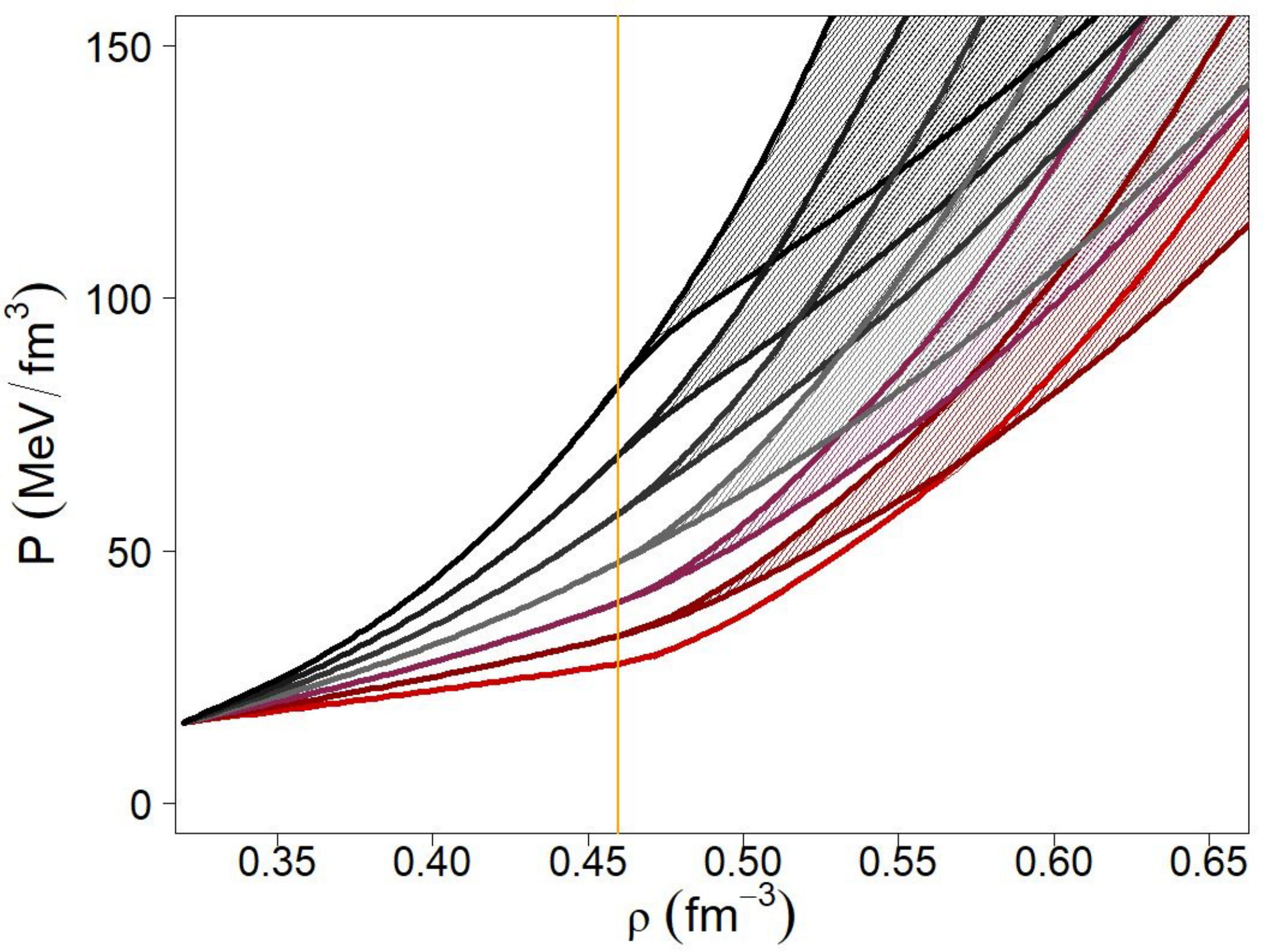}
\includegraphics[width=7.0cm]{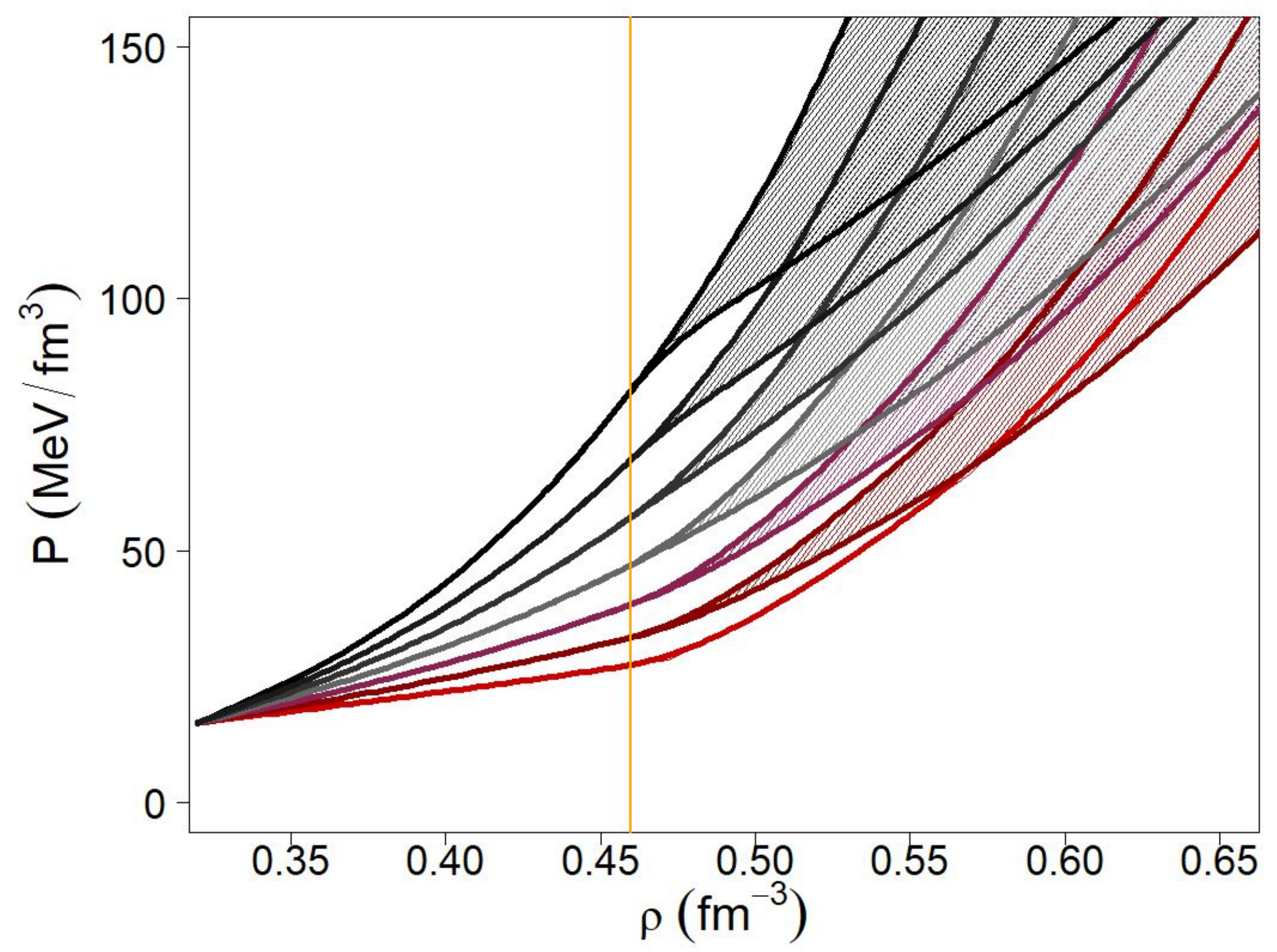}
\vspace*{0.5cm}
\caption{                                                                                                      
Spreading of the pressure from extension with polytropes as explained in the text.
The top, middle, and bottom rows show the results at N$^2$LO, N$^3$LO, and N$^4$LO, respectively.
Left and right: $\Lambda$=450 MeV and 500 MeV, respectively.
The vertical
coordinate axis and the vertical yellow line mark are located at
the two matching points. Only the combinations 
of polytropes which can support a maximum mass of at least 1.97 $M_{\odot}$ are retained. See text for more details.
} 
\label{pol}
\end{figure*}

\begin{figure*}[!t]
\includegraphics[width=8.0cm]{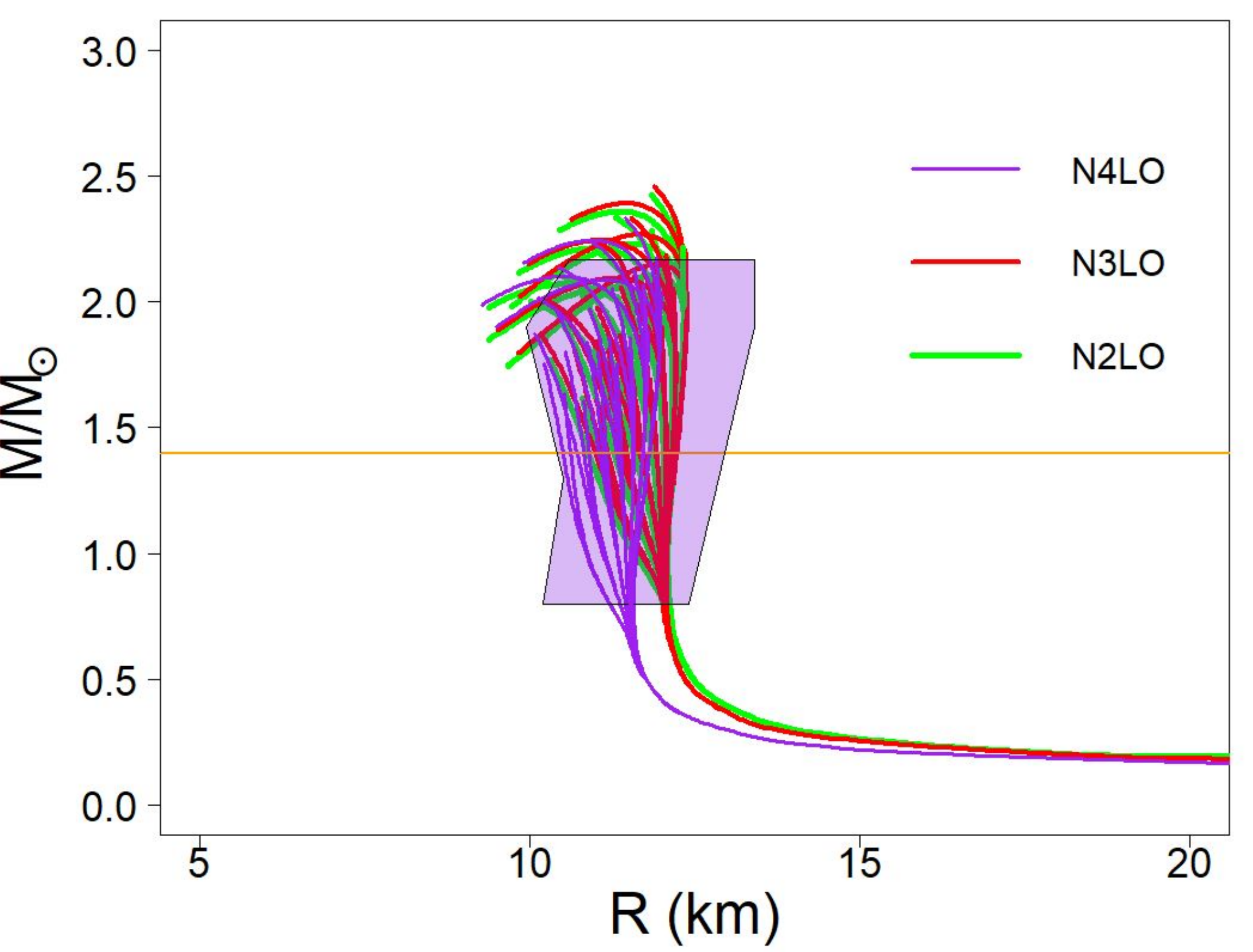}\hspace{.01in}
\includegraphics[width=8.0cm]{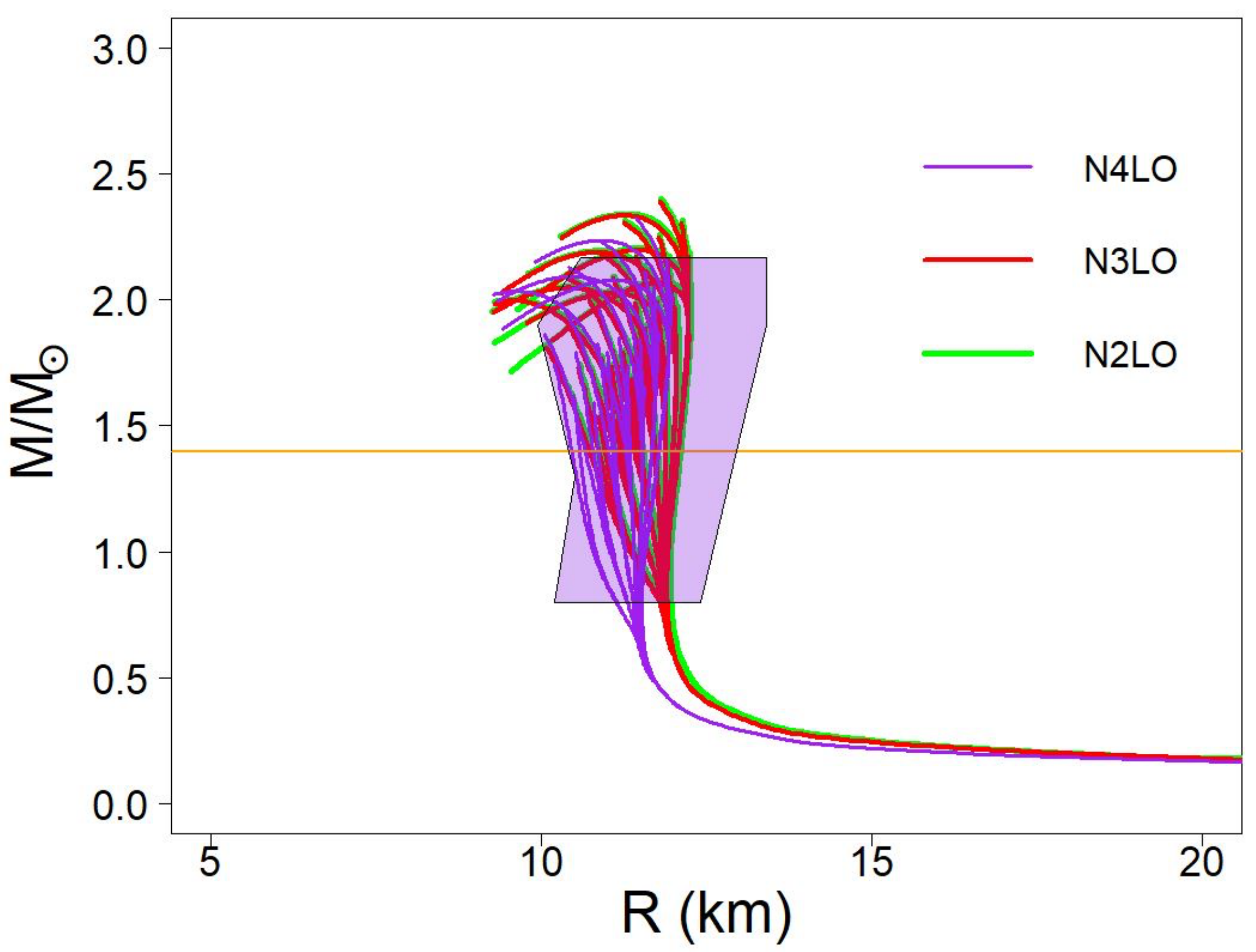}
\vspace*{0.7cm}
\caption{                                                                                                      
The mass {\it vs.} radius relation  for 
a neutron star at the indicated chiral order. Left: $\Lambda$=450 MeV; Right: $\Lambda$=500 MeV.
The various curves are obtained with the polytropic extension as explained in the text.                            
The purple curves are obtained extending the predictions at N$^4$LO, while the  
the red and the green curves are obtained extending the predictions at N$^3$LO and             
at N$^2$LO, respectively. 
The lavender shaded area is the constraint from Ref.~\cite{SLB13}.
} 
\label{polymass}
\end{figure*}

This procedure, and the corresponding spreading of the pressure, is demonstrated in 
Fig.~\ref{pol}.                                                    
Note that only combinations of $\Gamma _1$ 
and $\Gamma _2$ which 
can support 
a maximum mass of at least 1.97 $M_{\odot}$, are retained, to account for the observation of a pulsar 
with a mass of 2.01 $\pm$ 0.04 $M_{\odot}$~\cite{Ant13}.                                        
The $M(R)$ relations are shown in Fig.~\ref{polymass}, again for those polytropic extensions consistent with 
a maximum mass of at least 1.97 $M_{\odot}$.                                                            

Causality constraints impose limitations on the allowed 
values of $\Gamma_i$ and those are enforced in 
Fig.~\ref{polymass}.                                                        
 That is, one must require that the speed 
of sound in stellar matter is less than the speed of light. We recall that this condition can be expressed as 
$\frac{dP}{d\epsilon} <$ 1, where $\epsilon$ is the total energy density. It has been 
pointed out, however,                            
that the relation 
$v_s = c\sqrt{\frac{dP}{d\epsilon}}$ holds exactly if stellar matter is not dispersive (in
the presence of dispersion, the phase        
velocity of sound waves would not be well defined). Therefore, it is not entirely clear how strong a constraint
the above relations poses on the EoS~\cite{Weber}. 

We now proceed to estimate the value and the uncertainty for the radius of a 1.4 $M_{\odot}$ star.
To that end, we first average all values of the radius separately at N$^3$LO and N$^4$LO.
The error from the high-density continuation at 
N$^3$LO                                                          
is then combined in quadrature with 
the truncation error, estimated as the difference between the average radii at the two highest orders. 
More precisely, say that averaging over all polytropic solutions gives, at the two successive orders N$^3$LO and 
 N$^4$LO,       
\begin{equation}
R^{N3LO} = X ^{+\epsilon_u^{(3)}} _{-\epsilon_l^{(3)}} \; \; \mbox{and}  \; \; 
R^{N4LO} = Y ^{+\epsilon_u^{(4)}} _{-\epsilon_l^{(4)}} \; ,                  
\label{av}
\end{equation}
respectively. 
Then, we estimate the truncation error at N$^3$LO to be                          
\begin{equation}
\Delta = |X - Y|^{+\epsilon_u}_{-\epsilon_l} \; , 
\label{tr}
\end{equation}
where
\begin{equation}
\epsilon_u = \sqrt{\Big (\epsilon_u^{(3)}\Big)^2 +                   
\Big(\epsilon_u^{(4)}\Big)^2} \; ,                
\label{tu}
\end{equation}
and 
\begin{equation}
\epsilon_l = \sqrt{\Big(\epsilon_l^{(3)}\Big)^2 +                   
\Big(\epsilon_l^{(4)}\Big)^2} \; .                
\label{tl}
\end{equation}
Equations~(\ref{tu}-\ref{tl}) provide our estimate for 
the allowed range of the truncation error. For the radius of the 
1.4$M_{\odot}$ star, this procedure yields                                                            
\begin{equation}
 R^{N3LO} = (10.8 \; - \;12.8) \; \mbox{km} \;.
\label{r14}
\end{equation}

\begin{table}                
\centering
\caption{Adiabatic indices, 
$\Gamma_1$ and $\Gamma_2$, of the extension polytropes at the two matching densities, along with 
the radius and the central density of the 1.4 $M_{\odot}$ neutron star. The microscopic part of the 
predictions are obtained at N$^3$LO with $\Lambda$=450 MeV.} 
\begin{tabular}{c c c c }               
\hline
\hline
$\Gamma_1$ & $\Gamma_2$ & R (km) & $\rho$ ($fm^{-3}$)\\
\hline
\hline
1.5 & 3.5 & 10.98 & 0.684\\
1.5 & 4.0 & 11.09 & 0.640\\
1.5 & 4.5 & 11.17 & 0.612\\
    &     &       &      \\
2.0 & 3.5 & 11.36 & 0.611\\
2.0 & 4.0 & 11.42 & 0.585\\
2.0 & 4.5 & 11.45 & 0.567\\
    &     &       &      \\
2.5 & 3.0 & 11.66 & 0.567\\
2.5 & 3.5 & 11.68 & 0.548\\
2.5 & 4.0 & 11.70 & 0.536\\
2.5 & 4.5 & 11.71 & 0.528\\
    &     &       &      \\
3.0 & 3.0 & 11.92 & 0.501\\
3.0 & 3.5 & 11.92 & 0.498\\
3.0 & 4.0 & 11.92 & 0.496\\
3.0 & 4.5 & 11.92 & 0.494\\
    &     &       &      \\
3.5 & 2.5 & 12.07 & 0.457\\
3.5 & 3.0 & 12.06 & 0.457\\
3.5 & 3.5 & 12.06 & 0.456\\
3.5 & 4.0 & 12.06 & 0.456\\
3.5 & 4.5 & 12.06 & 0.455\\
    &     &       &      \\
4.0 & 2.0 & 12.16 & 0.429\\
4.0 & 2.5 & 12.16 & 0.429\\
4.0 & 3.0 & 12.16 & 0.429\\
4.0 & 3.5 & 12.16 & 0.429\\
4.0 & 4.0 & 12.16 & 0.429\\
4.0 & 4.5 & 12.16 & 0.429\\
    &     &       &      \\
4.5 & 1.5 & 12.23 & 0.411\\
4.5 & 2.0 & 12.23 & 0.411\\
4.5 & 2.5 & 12.22 & 0.411\\
4.5 & 3.0 & 12.22 & 0.411\\
4.5 & 3.5 & 12.22 & 0.411\\
4.5 & 4.0 & 12.22 & 0.411\\
4.5 & 4.5 & 12.22 & 0.411\\
\hline
\hline
\end{tabular}
\label{tab1} 
\end{table}

Table~\ref{tab1} shows the             
the radius and the central density of the 1.4 $M_{\odot}$ neutron star when the microscopic pressure 
is extended {\it via} polytropes with adiabatic indices as shown. 
The microscopic part of the 
predictions are obtained at N$^3$LO with $\Lambda$=450 MeV. (Very similar values are found with $\Lambda$=500 MeV.)
Clearly, the radius 
is nearly insensitive to the 
polytropic extension at 
the larger density, and changes only moderately 
due to variations of the first polytropic index between 1.5 and 4.5. 
In other words, 
the uncertainty reported in Eq.~(\ref{r14}) is relatively small given the huge uncertainty introduced in 
the pressure by the
polytropic continuation. Note that the central densities we predict for the average-mass star are 
typically in the order of, and can exceed 
3$\rho_0$, as can be seen from Table~\ref{tab1}. These densities are at or 
above the one marked by the yellow line in Fig.~\ref{pol}, where
the spreding of the pressure is as large as a factor of 4. Clearly, this 
indicates that the radius responds to pressures at much lower than central densities.
This is line with earlier observations~\cite{LP01} which found ``{\it...remarkable empirical correlation...
between the radii of 1 and 1.4 $M_{\odot}$ normal stars and the matter's pressure ...at fiducial densities of
1, 1.5 and 2$n_s$...}".
With our present observations, we wish to stress the point 
that the radius of the average-mass star    
bears the signature of the microscopic theory, being nearly insensitive to the phenomenological continuations.

Although masses of neutron stars can be and have been measured with high precision, simultaneous measurements
of radii are much more problematic. Some techniques do exist, such as those based on photospheric radius 
expansion~\cite{PRE}. Current observations have begun to determine the M(R) relation. In                          
Ref.~\cite{SLB13}, using data from both accreting sources and bursting sources, the authors determine 
the radius of a 1.4 M$_{\odot}$ neutron star to be between 10.4 and 12.9 km. Furthermore, from their 
Bayesian analysis of several EoS parametrized so as to be 
consistent with a baseline data set (see Ref.~\cite{SLB13} and references 
therein), they also determine the M(R) relation
for a range of neutron star masses. Our predictions are well within this constraint, shown in Fig.~\ref{polymass}  
 as the shaded purple area. Most recently, 
from LIGO/Virgo measurements the radius of a 1.4 $M_{\odot}$ neutron star was 
determined to be between 11.1 and 13.4 km~\cite{merg1,merg2}.

Before closing this section,
we take note of some other works aimed at incorporating 
aspects of chiral dynamics in the development of EoS suitable for astrophysical phenomena, 
such as Refs.~\cite{HS14,HLPS,Holt16}. In the 
latter reference, the authors calculate neutron star masses and radii 
with mean-field models whose parameters are made consistent with a chiral EoS at low to moderate densities. 
Constraints from chiral EFT on neutron star tidal deformabilities were investigated in Ref.~\cite{LH18}. 

\section{Summary and conclusions}                                  
\label{Conc} 

We calculated $M(R)$ relations for neutron star sequences using, as a starting point, EoS for neutron-rich
matter obtained in Brueckner-Hartree-Fock calculations based on modern high-quality chiral few-nucleon forces.
For densities above approximately twice normal density, we      
extrapolated the microscopic predictions using 
a family of polytropes whose range of adiabatic index is suggested by empirical analyses.                

With regard, specifically, to the radius of a 1.4 $M_{\odot}$ neutron star,
a central issue in this paper, we observe that our microscopic predictions up to 
about 2$\rho_0$, along with considerations of  theoretical error, essentially 
determine the radius of a 1.4 $M_{\odot}$ star. Our predictions fall between 10.8 and 12.8 km.                 

The simultaneous detection of gravitational and electromagnetic signals from the merger of two compact stars 
has recently begun a new era of ``multimessenger astronomy". In 
Ref.~\cite{merg2}, it is concluded with 90\% confidence 
from LIGO/Virgo measurements and otherwise very robust assumptions that the radius of a 1.4 $M_{\odot}$ is 
bound between 11.1 and 13.4 km.
Our chirally constrained predictions are in agreement with these constraints, which we find 
encouraging. 

 The quest for a unified microscopic approach to
neutron-rich systems, able to reach out to central densities of maximum-mass stars, remains an exceedingly 
complex task. Hopefully, a combination of theoretical, observational, and phenomenological efforts will
help us move towards a more complete picture of the EoS.

\section*{Acknowledgments}
This work was supported by 
the U.S. Department of Energy, Office of Science, Office of Basic Energy Sciences, under Award Number DE-FG02-03ER41270.


\begin{thebibliography} {100}  
\bibitem{merg1} B.P. Abbott {\it et al.} [LIGO Scientific and Virgo Collaborations], Phys. Rev. Lett.                 
{\bf 119}, 161101 (2017). 
\bibitem{merg2} Eemeli Annala, Tyler Gorda, Aleksi Kurkela, and Aleksi Vuorinen, arXiv:1711.02644 [astro-ph.HE]; and
references therein. 
\bibitem{ME11} R. Machleidt and D.R. Entem, Physics Report {\bf 503}, 1 (2011).    
\bibitem{EHM09} E. Epelbaum, H.-W. Hammer, and U.-G. Meissner, Rev. Mod. Phys. {\bf 81}, 1773 (2009).
\bibitem{Mac89} R. Machleidt, Adv. Nucl. Phys. {\bf 19}, 189 (1989).                                      
\bibitem{AV18} R.B. Wiringa, V.G.J. Stoks, and R. Schiavilla, Phys. Rev. C {\bf 51}, 38 (1995).          
\bibitem{poly} J.S. Read {\it et al.}, Phys. Rev. D {\bf 79}, 124032 (2009). 
\bibitem{tov} J.R. Oppenheimer and G.M. Volkoff, Phys. Rev. C {\bf 55}, 374 (1939). 

\bibitem{LP04} J.M. Lattimer and M. Prakash, Science {\bf 304}, 536 (2004). 
\bibitem{LP07} J.M. Lattimer and M. Prakash, Phys. Rep. {\bf 442}, 109 (2007). 
\bibitem{LP16} J.M. Lattimer and M. Prakash, Phys. Rep. {\bf 621}, 127 (2016). 
\bibitem{SLB13} A.W. Steiner, J.M. Lattimer, and E.F. Brown, Astrophys. J. {\bf 765}, L5 (2013).
\bibitem{Oz13} F. {\"O}zel, Rep. Prog. Phys. {\bf 76}, 016901 (2013).
%
\bibitem{NLO} E. Marji, A. Canul, Q. MacPherson, R. Winzer, Ch. Zeoli, D.R. Entem, and R. Machleidt, Phys. Rev. C {\bf 88}, 054002 (2013). 
\bibitem{EM03} D.R. Entem and R. Machleidt, Phys. Rev. C {\bf 68}, 041001 (2003).

\bibitem{EGM05} E. Epelbaum, W. Gl{\" o}kle, and U.-G. Meissner, Nucl. Phys. A {\bf 747}, 362 (2005).
\bibitem{EKM15} E. Epelbaum, H. Krebs, and U.-G. Meissner, Phys. Rev. Lett. {\bf 115}, 122301 (2015).

\bibitem{n4lo} D.R. Entem, R. Machleidt, and Y. Nosyk, Phys. Rev. C {\bf 96}, 024004 (2017).    


 \bibitem{Epe02} 
E. Epelbaum, A. Nogga, W. Gl\"ockle, H. Kamada, U.-G. Mei\ss ner, and H. Witala,
{Phys. Rev. C} {\bf 66}, 064001 (2002).
     \bibitem{NRQ10} 
  P.~Navratil, R.~Roth, and S.~Quaglioni,
{Phys.\ Rev.\ C} {\bf 82}, 034609 (2010).
\bibitem{Viv13}
M. Viviani, L. Girlanda, A. Kievsky, and L. E. Marcucci, 
{Phys. Rev. Lett.} {\bf 111}, 172302 (2013).
  \bibitem{Gol14}
J.~Golak {\it et al.},
{Eur.\ Phys.\ J.\ A} {\bf 50}, 177 (2014).
\bibitem{Kal12} 
  N.~Kalantar-Nayestanaki, E.~Epelbaum, J.~G.~Messchendorp, and A.~Nogga,
  Rept.\ Prog.\ Phys.\  {\bf 75}, 016301 (2012).
\bibitem{Nav16} 
  P.~Navratil, S.~Quaglioni, G.~Hupin, C.~Romero-Redondo, and A.~Calci,
  Phys.\ Scripta {\bf 91}, 053002 (2016).
\bibitem{Coraggio07}
L. Coraggio, A. Covello, A. Gargano, N. Itaco, T. T. S. Kuo,
D. R. Entem, and R. Machleidt, Phys. Rev. C {\bf 75}, 024311 (2007).
\bibitem{Coraggio10}
L. Coraggio, A. Covello, A. Gargano, and N. Itaco, Phys. Rev. C {\bf
  81}, 064303 (2010).
\bibitem{Coraggio12}
L. Coraggio, A. Covello, A. Gargano, N. Itaco, and T. T. S. Kuo,
Ann. Phys. {\bf 327}, 2125 (2012).
 \bibitem{Hag12a}
H. Hagen, M. Hjorth-Jensen, G. R. Jansen, R.  Machleidt, and T. Papenbrock,
{Phys. Rev. Lett.} {\bf 108}, 242501 (2012).
\bibitem{Hag12b}
H. Hagen, M. Hjorth-Jensen, G. R. Jansen, R.  Machleidt, and T. Papenbrock,
{Phys. Rev. Lett.} {\bf 109}, 032502 (2012).
 \bibitem{BNV13} 
  B.~R.~Barrett, P.~Navratil, and J.~P.~Vary,
{Prog.\ Part.\ Nucl.\ Phys.}  {\bf 69}, 131 (2013).
\bibitem{Gez13}
A.\ Gezerlis, I.\ Tews, E.\ Epelbaum, S.\ Gandolfi, K.\ Hebeler, A.\ Nogga, and A.\ Schwenk, 
Phys.\ Rev.\ Lett.\ 111, 032501 (2013).
  \bibitem{Her13} 
  H.~Hergert, S.~K.~Bogner, S.~Binder, A.~Calci, J.~Langhammer, R.~Roth, and A.~Schwenk,
{Phys.\ Rev.\ C} {\bf 87}, 034307 (2013).
  \bibitem{Hag14a} 
  G.~Hagen, T.~Papenbrock, M.~Hjorth-Jensen, and D.~J.~Dean,
{Rept.\ Prog.\ Phys.}  {\bf 77}, 096302 (2014).
\bibitem{Som14} 
  V.~Som\`{a}, A. Cipollone, C.~Barbieri, P. Navratil, and T. Duget,
{Phys.\ Rev.\ C}  {\bf 89}, 061301(R) (2014).
\bibitem{Heb15}
K. Hebeler, J. D. Holt, J. Men\'{e}ndez, and A. Schwenk,
Ann. Rev. Nucl. Part. Sci. {\bf 65}, 457 (2015).
\bibitem{Hag16}
G.~Hagen {\it et al.},
  Nature Phys.\  {\bf 12}, 186 (2015).
  \bibitem{Car15}
  J.~Carlson, S.~Gandolfi, F.~Pederiva, S.~C.~Pieper, R.~Schiavilla, K.~E.~Schmidt, and R.~B.~Wiringa,
{Rev.\ Mod.\ Phys.}  {\bf 87}, 1067 (2015).
  \bibitem{Her16} 
  H.~Hergert, S.~K.~Bogner, T. D. Morris, A.~Schwenk, and K. Tsukiyama,
{Phys.\ Rep.} {\bf 621}, 165 (2016).
\bibitem{Hol17}
  J.\ W.\ Holt and N.\ Kaiser, Phys.\ Rev.\ C {\bf 95}, 034326 (2017).
  \bibitem{Sim17} 
  J.~Simonis, S.~R.~Stroberg, K.~Hebeler, J.~D.~Holt, and A.~Schwenk,
{Phys. Rev. C} {\bf 96}, 014303 (2017)
 \bibitem{Mor17} 
T. D. Morris,  J.~Simonis, S.~R.~Stroberg, C. Stumpf, G. Hagen, J. D. Holt, 
G. R. Jansen, T. Papenbrock, R. Roth, and A.~Schwenk,
  arXiv:1709.02786 [nucl-th].

\bibitem{HS10} 
  K.~Hebeler and A.~Schwenk,
{Phys.\ Rev.\ C} {\bf 82}, 014314 (2010).
  \bibitem{Heb11} 
  K.~Hebeler, S. K. Bogner, R. J. Furnstahl, A. Nogga, and A.~Schwenk,
{Phys.\ Rev.\ C} {\bf 83}, 031301(R) (2011).
\bibitem{Baa13}
G.~Baardsen, A.~Ekstr\"{o}m, G.~Hagen and M.~Hjorth-Jensen,
  Phys.\ Rev.\ C {\bf 88}, 054312 (2013)
\bibitem{Hag14b} 
  G.~Hagen, T.~Papenbrock, A.~Ekstr\"{o}m, K.~A.~Wendt, G.~Baardsen, S.~Gandolfi, M.~Hjorth-Jensen, and C.~J.~Horowitz,
{Phys.\ Rev.\ C} {\bf 89}, 014319 (2014).
  \bibitem{Cor13}
L. Coraggio, J. W. Holt, N. Itaco, R. Machleidt, and F. Sammarruca,  
{Phys. Rev. C} {\bf 87}, 014322 (2013).
 \bibitem{Cor14}
  L. Coraggio, J. W. Holt, N. Itaco, R. Machleidt, L. E. Marcucci, and F. Sammarruca,  
{Phys. Rev. C} {\bf 89}, 044321 (2014).
\bibitem{Sam15}
  F. Sammarruca, L. Coraggio, J. W. Holt, N. Itaco, R. Machleidt, and L. E. Marcucci,   
{Phys. Rev. C} {\bf 91}, 054311 (2015).
\bibitem{Dri16} 
  C.~Drischler, A.~Carbone, K.~Hebeler, and A.~Schwenk,
  Phys.\ Rev.\ C {\bf 94}, 054307 (2016).
\bibitem{Tew16} 
  I.~Tews, S.~Gandolfi, A.~Gezerlis, and A.~Schwenk,
  Phys.\ Rev.\ C {\bf 93}, 024305 (2016).

\bibitem{Wel14}
  C.~Wellenhofer, J.~W.~Holt, N.~Kaiser, and W.~Weise, Phys.\ Rev.\ C {\bf 89}, 064009 (2014).
\bibitem{Wel15}
  C.~Wellenhofer, J.~W.~Holt, and N.~Kaiser, Phys.\ Rev.\ C {\bf 92}, 015801 (2015).

\bibitem{Bac09}
  S.\ Bacca, K.\ Hally, C.\ J.\ Pethick, and A.\ Schwenk, Phys. Rev. C {\bf 80}, 032802 (2009).
\bibitem{Bar14}
  A.\ Bartl, C.\ J.\ Pethick, and A.\ Schwenk, Phys.\ Rev.\ Lett.\ {\bf 113} 081101 (2014).
\bibitem{Rap15}
  E.\ Rrapaj, J.\ W.\ Holt, A.\ Bartl, S.\ Reddy, and A.\ Schwenk, Phys.\ Rev.\ C {\bf 91}, 035806 (2015).
\bibitem{Bur16}
  M.\ Buraczynski and A.\ Gezerlis, Phys. Rev. Lett. {\bf 116}, 152501 (2016).
\bibitem{Hol16}
  J.\ W.\ Holt, N.\ Kaiser, and G.\ A.\ Miller, Phys.\ Rev.\ C {\bf 93}, 064603 (2016).
\bibitem{Bir17}
  J.\ Birkhan et al., Phys.\ Rev.\ Lett.\ {\bf 118}, 252501 (2017).
\bibitem{Rot17}
  J.\ Rotureau, P.\ Danielewicz, G.\ Hagen, F.\ Nunes, and T.\ Papenbrock, Phys.\ Rev.\ C {\bf 95},024315 (2017).

\bibitem{Hofe} M. Hoferichter {\it et al.}, Phys. Rev. Lett. {\bf 115}, 192301 (2015); Phys. Rep. {\bf 625}, 1 (2016).
\bibitem{FS+18} F. Sammarruca, L.E. Marcucci, L. Coraggio, J.W. Holt, N. Itaco, and R. Machleidt, arXiv: 1807.06640.

\bibitem{HKW} J.W. Holt, N. Kaiser, and W. Weise, Phys. Rev. C {\bf 79}, 054331 (2009); 
Phys. Rev. C {\bf 81}, 024002 (2010). 

\bibitem{Marc18} L. E. Marcucci, A. Kievsky, S. Rosati,
  R. Schiavilla, and M. Viviani,
  Phys. Rev. Lett. {\bf 108}, 052502, (2012);
  Erratum: accepted for publication in Phys. Rev. Lett. (2018).


\bibitem{KGE12}
H. Krebs, A. Gasparyan, and E. Epelbaum, 
Phys. Rev. C {\bf 85}, 054006 (2012).


\bibitem{FM57}
J.-I. Fujita and H. Miyazawa, Prog. Theor. Phys. {\bf 17}, 360 (1957). 

\bibitem{HT} Michael I. Haftel and Frank Tabakin, Nucl. Phys.   
 {\bf A158}, 1 (1970).      

\bibitem{HW} B.K. Harrison and J.A. Wheeler in {\it Gravitation Theory and Gravitational Collapse}  
(B.K. Harrison, K.S. Thorne, M. Wakano, and J.A. Wheeler, eds.), pp.~1-177, University of Chicago Press, 
Chicago (1965). 
\bibitem{NV} J.W. Negele and D. Vautherin, {\it Nucl. Phys.} {\bf A178}, 123 (1973).                
\bibitem{Ant13} J. Antoniadis {\it et al.}, Science {\bf 340}, 6131 (2013). 
\bibitem{Weber} Fridolin Weber, in ``Pulsar as Astrophysical Laboratories for Nuclear and Particle Physics",
IOP, Bristol and Philadelphia, 1999.
\bibitem{LP01} J.M. Lattimer and M. Prakash, Ap. J. {\bf 550}, 426 (2001). 
\bibitem{PRE} J. van Paradijs, Astrophys. J. {\bf 234}, 609 (1079). 
\bibitem{HLPS} K. Hebeler, J.M. Lattimer, C.J. Pethick, A. Schwenk, Phys. Rev. Lett. {\bf 105}, 161102 (2010).
\bibitem{HS14} K. Hebeler and A. Schwenk, Eur. Phys. J. A {\bf 50}, 11 (2014). 
\bibitem{Holt16} Ermal Rrapaj, Alessandro Roggero, and Jeremy W. Holt, 
Phys. Rev. C {\bf 93}, 065801 (2016). 
\bibitem{LH18} Yeunhwan Lim and Jeremy W. Holt, arXiv:1803.02803v1.    
%
%
\end{thebibliography}
\end{document}